\documentclass[aps,prl,groupedaddress,twocolumn,superscriptaddress]{revtex4-1}
\usepackage{graphicx}
\usepackage{xfrac}
\usepackage{amssymb}
\usepackage{verbatim}

\usepackage{natbib}
\begin{document}

\title{Towards Neutron Capture on Exotic Nuclei: \\ Demonstrating $(d,p\gamma)$ as a Surrogate Reaction for $(n,\gamma)$}

\author{A. Ratkiewicz}
\email[to whom correspondence should be addressed at: ]{ratkiewicz1@llnl.gov}
\affiliation{Lawrence Livermore National Laboratory, Livermore, California 94550, USA}
\affiliation{Department of Physics and Astronomy, Rutgers University, New Brunswick, New Jersey 08901, USA}

\author{J.A. Cizewski}
\affiliation{Department of Physics and Astronomy, Rutgers University, New Brunswick, New Jersey 08901, USA}

\author{J.E. Escher}
\affiliation{Lawrence Livermore National Laboratory, Livermore, California 94550, USA}

\author{G. Potel}
\affiliation{Michigan State University, East Lansing, Michigan 48824, USA}
\affiliation{Facility for Rare Isotope Beams, East Lansing, Michigan 48824, USA}

\author{J.T. Burke}
\affiliation{Lawrence Livermore National Laboratory, Livermore, California 94550, USA}

\author{R.J. Casperson}
\affiliation{Lawrence Livermore National Laboratory, Livermore, California 94550, USA}

\author{M. McCleskey}
\affiliation{Cyclotron Institute, Texas A\&M University, College Station, Texas 77843, USA}

\author{R.A.E. Austin}
\affiliation{Astronomy and Physics Department, Saint Mary's University, Halifax, Nova Soctia BH3 3C3, Canada}

\author{S. Burcher}
\affiliation{Department of Physics and Astronomy, Rutgers University, New Brunswick, New Jersey 08901, USA}

\author{R.O. Hughes}
\affiliation{Lawrence Livermore National Laboratory, Livermore, California 94550, USA}
\affiliation{Department of Physics, University of Richmond, Virginia 23173, USA}

\author{B. Manning}
\affiliation{Department of Physics and Astronomy, Rutgers University, New Brunswick, New Jersey 08901, USA}

\author{S.D. Pain}
\affiliation{Physics Division, Oak Ridge National Laboratory, Oak Ridge, Tennessee 37831, USA}

\author{W.A. Peters}
\affiliation{Oak Ridge Associated Universities, Oak Ridge, Tennessee 37831, USA}

\author{S. Rice}
\affiliation{Department of Physics and Astronomy, Rutgers University, New Brunswick, New Jersey 08901, USA}

\author{T.J. Ross}
\affiliation{Department of Physics, University of Richmond, Virigina 23173, USA}

\author{N.D. Scielzo}
\affiliation{Lawrence Livermore National Laboratory, Livermore, California 94550, USA}

\author{C. Shand}
\affiliation{Department of Physics and Astronomy, Rutgers University, New Brunswick, New Jersey 08901, USA}
\affiliation{Department of Physics, University of Surrey, Guildford, Surrey, GU2 7XH, UK}

\author{K. Smith}
\affiliation{Los Alamos National Laboratory, Los Alamos, New Mexico 87544, USA}

\date{\today}

\begin{abstract}
The neutron-capture reaction plays a critical role in the synthesis of the elements in stars and is important for societal applications including nuclear power generation and stockpile-stewardship science. However, it is difficult --- if not impossible --- to directly measure neutron capture cross sections for the exotic, short-lived nuclei that participate in these processes. In this Letter we demonstrate a new technique which can be used to indirectly determine neutron-capture cross sections for exotic systems. This technique makes use of the $(d,p)$ transfer reaction, which has long been used as a tool to study the structure of nuclei. Recent advances in reaction theory, together with data collected using this reaction, enable the determination of neutron-capture cross sections for short-lived nuclei. A benchmark study of the $^{95}$Mo$(d,p)$ reaction is presented, which illustrates the approach and provides guidance for future applications of the method with short-lived isotopes produced at rare isotope accelerators.
\end{abstract}


\maketitle
Essentially all of the heavy elements are synthesized in astrophysical environments by processes that involve neutron capture. The slow neutron-capture process (the \textit{s} process) occurs predominantly in the low neutron flux in AGB stars, yielding a nucleosynthesis path that typically deviates only one or two neutrons from $\beta$-stability. In contrast, the rapid neutron-capture process (the \textit{r} process) involves exotic neutron-rich nuclei and requires explosive stellar scenarios with high neutron fluences. The \textit{r} process is responsible for the creation of roughly half of the elements between iron and bismuth and synthesizes heavy nuclei through the rapid production of neutron-rich nuclei via neutron capture and subsequent $\beta$ decay.

The recent observation of the gravitational waves associated with a neutron-star merger~\cite{PhysRevLett.119.161101}, and the subsequent kilonova understood to be powered by the decay of lanthanides~\cite{2041-8205-848-2-L27,naturePian2017}, demonstrated that neutron-star mergers are an important \textit{r}-process site, especially for the heaviest elements. However, \textit{r}-process abundance patterns are sensitive to astrophysical conditions (cf.~\cite{Mumpower201686}). In a ``cold'' \textit{r} process (which could occur in a neutron star merger or with the highly accelerated neutrino-driven winds following a core-collapse supernova), equilibrium between neutron capture $(n,\gamma)$ and photo-dissociation $(\gamma,n)$ rapidly breaks down, so the rate at which neutron capture proceeds will affect the final \textit{r}-process abundance pattern. The timescales of the cold \textit{r} process are such that competition between neutron capture and $\beta$ decay occurs during the bulk of the \textit{r}-process nucleosynthesis. Neutron-capture rates on unstable nuclei affect the final observed abundance patterns even in the traditional ``hot'' \textit{r} process (thought to occur in the neutrino-driven winds in a proto-neutron star resulting from a core-collapse supernova) during the eventual freeze-out, when $(n,\gamma)\leftrightharpoons(\gamma,n)$ equilibrium no longer occurs. Accordingly, neutron capture is influential in determining the final \textit{r}-process abundance pattern, especially beyond $(n,\gamma)\leftrightharpoons(\gamma,n)$ equilibrium. Therefore, measuring $(n,\gamma)$ rates on key neutron-rich nuclei continues to be an important component in understanding \textit{r}-process abundance patterns and constraining the astrophysical sites for \textit{r}-process nucleosynthesis as a function of mass~\cite{ARCONES20171}.

Cross sections for neutron capture on the nuclei that participate in the \textit{r} process have proven to be difficult to determine because neither the heavy nucleus nor the neutron can serve as a target due (in part) to their short lifetimes. Of the nuclear data used to predict the final abundance pattern produced by \textit{r}-process nucleosynthesis, the $(n,\gamma)$ rates are the most poorly constrained. Experimental constraints on these rates are necessary, as the final abundance pattern calculated by global nucleosynthesis models is sensitive to them~\cite{Mumpower201686,PhysRevC.79.045809,doi:10.1063/1.4867191}. Uncertainties in $(n,\gamma)$ cross sections on unstable nuclei also impact applications in nuclear energy, nuclear forensics, and stockpile-stewardship science (cf.~\cite{CARLSON201768}). A number of indirect methods for extracting or constraining these critically-important cross sections have been developed in recent years, among them the ``Beta-Oslo'' method ~\cite{PhysRevLett.113.232502} and the $\gamma$-ray strength function method \cite{PhysRevC.82.064610}. 

In this Letter, we demonstrate that the \textit{Surrogate Reactions Method (SRM)}~\cite{RevModPhys.84.353}, which was first introduced to infer cross sections for neutron-induced fission~\cite{nseBritt1979}, can be used to determine $(n,\gamma)$ cross sections using experimentally-accessible charged-particle reactions. The method uses the fact that the neutron-induced reaction of interest proceeds through the formation of an intermediate compound nucleus (CN), which subsequently decays. In a surrogate-reaction experiment, a target-projectile pair is chosen to form a CN that has the same excitation energy and neutron and proton numbers as the one produced in the desired (neutron-induced) reaction. 

The $(d,p)$ reaction has long been used as a probe of single-particle nuclear structure. Most of these measurements were conducted in normal kinematics; a deuteron beam was used to interrogate a stable or long-lived target. However, the outlook for $(d,p)$ reaction measurements with exotic nuclei in inverse kinematics, in which a radioactive beam is impinged on a deuterated target, is bright. Radioactive ion accelerators are beginning to be able to provide high-intensity beams at rates $\geq10^{4}$ particles per second, necessary for these measurements. In addition, $(d,p)$ in inverse kinematics yields a relatively clean measurement in which the only reaction particles at back angles in the laboratory are $(d,p)$ protons; particles from (in)elastic scattering and other reaction channels are emitted at angles forward of $90^{\circ}$ in the laboratory. 

Calculations of the $(d,p)$ reaction can be made more tractable by approximating the many-body $d+A$ system as a three-body problem: $p+n+A$. Deuterons are weakly bound; when a deuteron encounters a massive nucleus ($A$), it can either undergo \textit{elastic breakup} (EB), in which $A$ is left in its ground state ($A(d,pn)A$), or \textit{non-elastic breakup} (NEB), which has several possible exit channels and proceeds as $A(d,p)X$~\cite{PhysRevC.92.034611}. Interest in developing a robust description of the $(d,p)$ reaction has recently been renewed, in part by the promise this reaction shows as an $(n,\gamma)$ surrogate reaction (cf. ~\cite{PhysRevC.92.034611,Potel2017} and references therein). 

Before the SRM can be used to deduce $(n,\gamma)$ cross sections on unstable nuclei, the technique must be validated in normal kinematics against known cross sections. The $^{95}$Mo target was chosen to validate this method for surrogate reaction analysis because the $(n,\gamma)$ cross section has been directly measured at neutron energies up to 200 keV~\cite{DELMUSGROVE1976108} and the level scheme of $^{96}$Mo is established up to relatively high excitation energy~\cite{Abriola20082501}. Modeling the $\gamma$ cascade is simplified if the final nucleus is even-even, with a strong $2^{+}_{1}\rightarrow0^{+}_{g.s.}$ transition collecting most of the $\gamma$-ray strength depopulating higher-lying excited levels populated by the surrogate reaction, as is the case in $^{96}$Mo.

Early applications of the surrogate approach employed the ``Weisskopf-Ewing'' (WE) approximation, in which the decay of a CN is considered to be independent of its spin and parity, i.e. it is assumed that the decay of the CN is identical in both cases. This approximation has been successfully used to indirectly determine $(n,f)$ cross sections (see, e.g.~\cite{PhysRevC.67.024610,PhysRevC.68.034610,PhysRevC.74.054601,Ressler:2011zz,KESSEDJIAN2010297,PhysRevC.81.034608,PhysRevC.85.024613,PhysRevC.90.014304,PhysRevC.90.034601}). However, attempts to extract an ($n,\gamma)$ cross section from surrogate-reaction data through the WE approximation resulted in large disagreements between the extracted and known cross sections~\cite{PhysRevC.81.034608,BOUTOUX2012319,PhysRevC.81.011602}. This has been attributed to differences in the angular momentum (spin and parity) distributions with which compound nuclei are formed in the desired $(n,\gamma)$ and surrogate reactions (see, e.g.~\cite{PhysRevC.75.055807,RevModPhys.84.353,PhysRevC.81.024612,PhysRevC.92.034611}). Such differences would strongly influence the decay of the CN, which invalidates the premise of the WE approximation.

In the present Letter, we move beyond the WE approximation by utilizing a recently-developed $(d,p)$ reaction description to account for the spin-parity mismatch and its effects on the observed CN decay. This new description of the $(d,p)$ reaction~\cite{PhysRevC.92.034611,Potel2017} allows the reaction channel forming the desired CN to be selected and the formation of the CN, including the spins and parties of the states populated, to be calculated as a function of excitation energy. Therefore, the predicted formation and subsequent decay can be used to connect the experimentally-measured surrogate decay to the cross-section calculations that need to be constrained. In this Letter we demonstrate that an accurate neutron-capture cross section can be obtained from surrogate-reaction data using this method.

The largest source of uncertainty in neutron-capture calculations arises from insufficient knowledge of the nuclear-structure properties that enter the description of the CN decay. Specifically, nuclear level densities (NLD) and $\gamma$-ray strength functions ($\gamma$SF) determine whether the CN decays primarily by neutron or $\gamma$-ray emission. A newly-developed method by Escher and colleagues~\cite{refId0,escherTR738195} constrains the parameters in the decay models via Bayesian fits to the experimentally-measured surrogate coincidence probabilities. The constrained parameters are subsequently used to calculate the cross section for the $(n,\gamma)$ reaction. 

\begin{figure}
\includegraphics[width=8cm,clip]{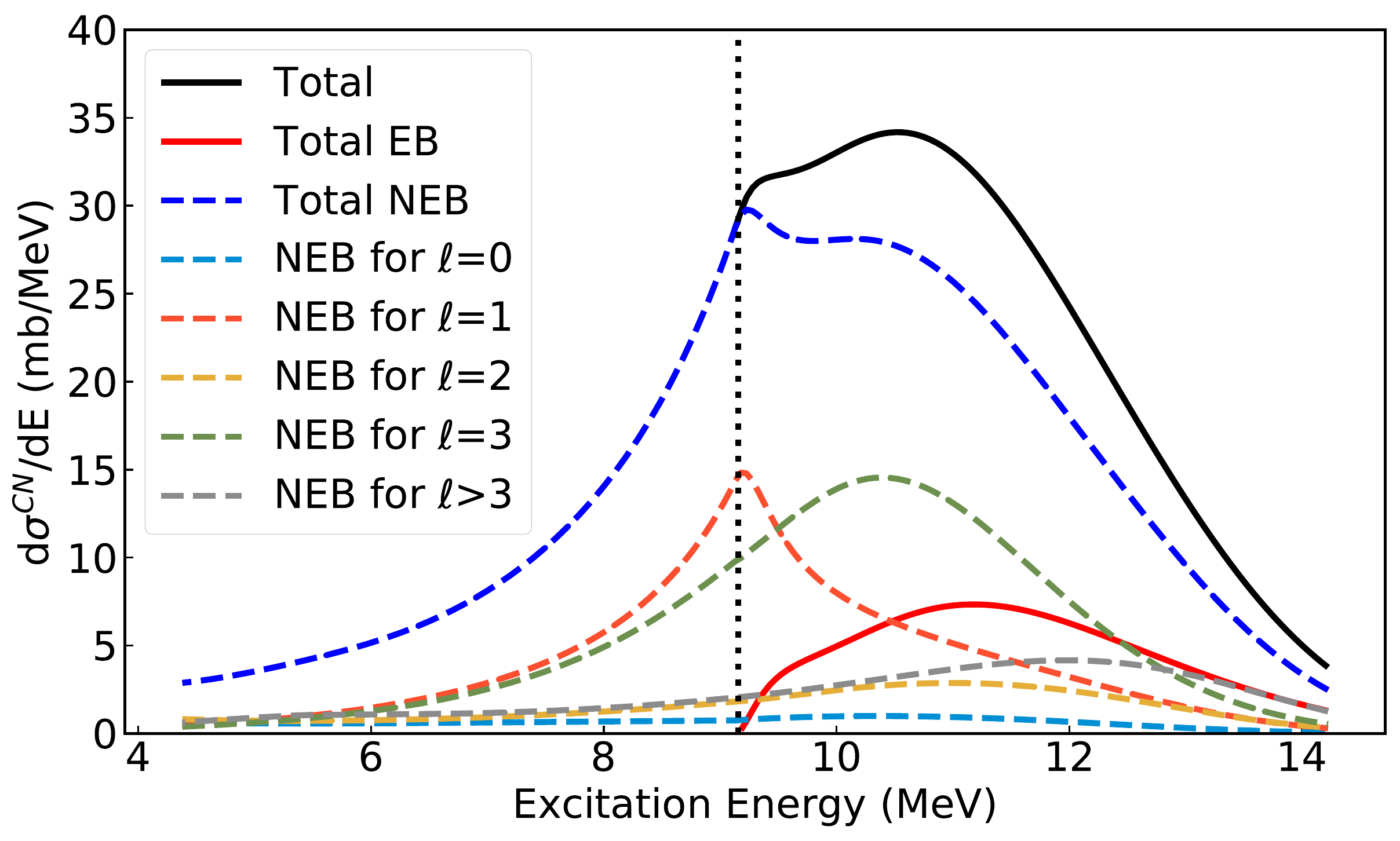}
	\caption{(Color online). Calculations of the $^{95}$Mo$(d,p)$ cross section as a function of excitation energy decomposed into total elastic breakup (EB, red line) and nonelastic breakup (NEB, dashed lines) components. The NEB component is further decomposed into contributions with different orbital angular momenta of the captured neutron. The vertical dotted line corresponds to the $S_{N}$ in the $^{96}$Mo CN. These calculations are integrated over the experimental center-of-mass angular range of 29$^{\circ}$--59$^{\circ}$. \label{fig:greg_calc}}
\end{figure}

\begin{figure}
\includegraphics[scale=0.48,clip]{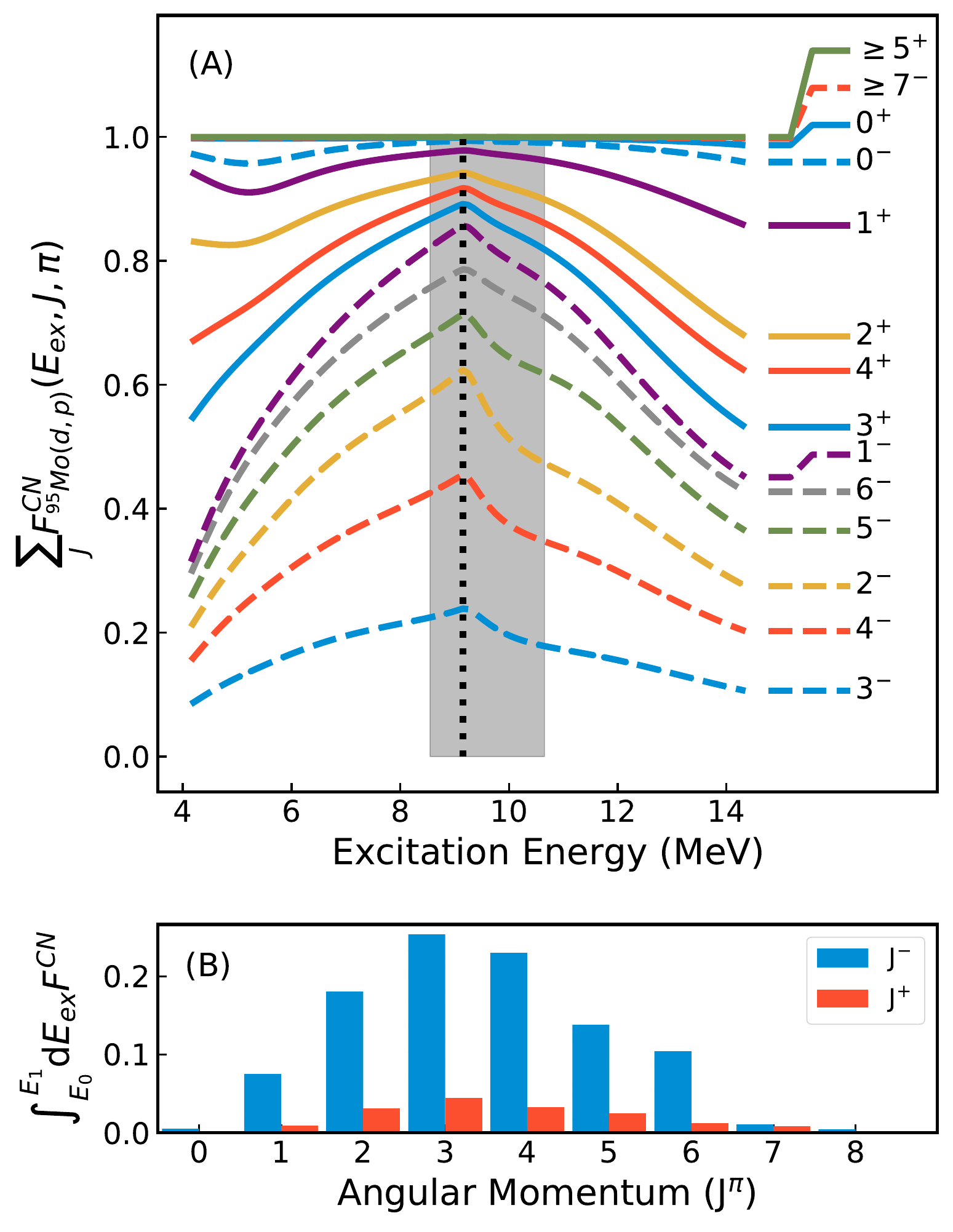}
	\caption{(Color online). (A) Calculations of the cumulative probability of forming the CN through $^{95}$Mo$(d,p)$ ($\sum F^{CN}_{{}^{95}Mo(d,p)}$). The shaded region, from $E_{0} = 8.55$ MeV to $E_{1} = 10.65$ MeV, indicates the excitation energies over which the surrogate data are fit. The states are plotted from largest contribution over the fitting range ($3^{-}$) to least ($\geq5^{+}$). The vertical dotted line represents $S_{N}$. (B) Histogram of the total contribution to CN formation over the shaded range in (A) as a function of angular momentum, decomposed into positive and negative parities, and normalized to one over the integration region. Negative-parity, low-J states dominate near $S_{N}$.\label{fig:abb_spin_dist}}
\end{figure}

The neutron-capture reaction cross section can be expressed in the Hauser-Feshbach formalism as~\cite{RevModPhys.84.353, refId0}:
\begin{equation}\label{eqn:HF_ng_CX}
\sigma_{\alpha\chi}\left(E_{n}\right) = \sum_{J,\pi} \sigma_{\alpha}^{CN}\left(E_{ex},J,\pi\right)G_{\chi}^{CN}\left(E_{ex},J,\pi\right).
\end{equation}
Here, $\sigma^{CN}_{\alpha}(E_{ex},J,\pi)$ is the cross section for forming a CN with some excitation energy $E_{ex}$ and spin-parity $J^{\pi}$ through the entrance channel $\alpha = n + {}^{A}Z$. The individual $\sigma^{CN}(E_{ex},J,\pi)$ can be calculated with an appropriate neutron-nucleus optical potential, such as that described in~\cite{KONING2003231}. However, the branching ratios $G_{\chi}^{CN}(E_{ex},J,\pi)$ for the decay of the CN through the exit channel $\chi$ (here $\gamma$-ray emission), depend on uncertain structural properties of the nucleus, in particular upon the NLD and $\gamma$SF, and thus need to be constrained. This is done with the aid of surrogate reaction data. The probability of forming the CN ($^{A+1}Z$) through a surrogate reaction through the entrance channel $\delta = d + {}^{A}Z$ and subsequently decaying through the exit channel of interest, $\chi = p + {}^{A+1}Z$, is given by:
\begin{equation}\label{eqn:full_g_prob}
P_{\delta\chi}(E_{ex},\theta_{p}) = \sum_{J,\pi} F_{\delta}^{CN}\left(E_{ex},J,\pi,\theta_{p}\right)G_{\chi}^{CN}\left(E_{ex},J,\pi\right).
\end{equation}
$\theta_{p}$ represents the angle between the outgoing proton and the beam axis. $F_{\delta}^{CN}\left(E_{ex},J,\pi,\theta_{p}\right)$ is the probability of forming the CN in the surrogate reaction and is determined by treating the deuteron-induced reaction as a two-step process~\cite{PhysRevC.92.034611,Potel2017}: in the first step the deuteron breaks up, releasing the neutron. The second step describes the interaction of the neutron with the target nucleus. The reaction cross section can then be decomposed into components due to EB and NEB (which includes neutron capture). The fusion of the $d+A$ system and subsequent evaporation of a proton is not included in these calculations. In the energy region of interest (near $S_{N}$) contributions from this process are expected to be very small, based on the analysis in ~\cite{Potel2017}. The NEB component is then further decomposed by the transfer of angular momentum (see Fig.~\ref{fig:greg_calc}), which gives the CN entry spin-parity distribution $F_{\delta}^{CN}\left(E_{ex},J,\pi\right)$. The single-particle structure of the CN strongly affects its spin-parity distribution, as shown in~\cite{ANDERSEN197033}. This dependence is included in the description of the neutron-target interaction. For the $^{95}$Mo$(d,p\gamma)$ reaction the $F_{\delta}^{CN}\left(E_{ex},J,\pi\right)$ are shown in Fig.~\ref{fig:abb_spin_dist} as a function of the excitation energy in the $^{96}$Mo CN.

\begin{figure}
\includegraphics[width=8cm,clip]{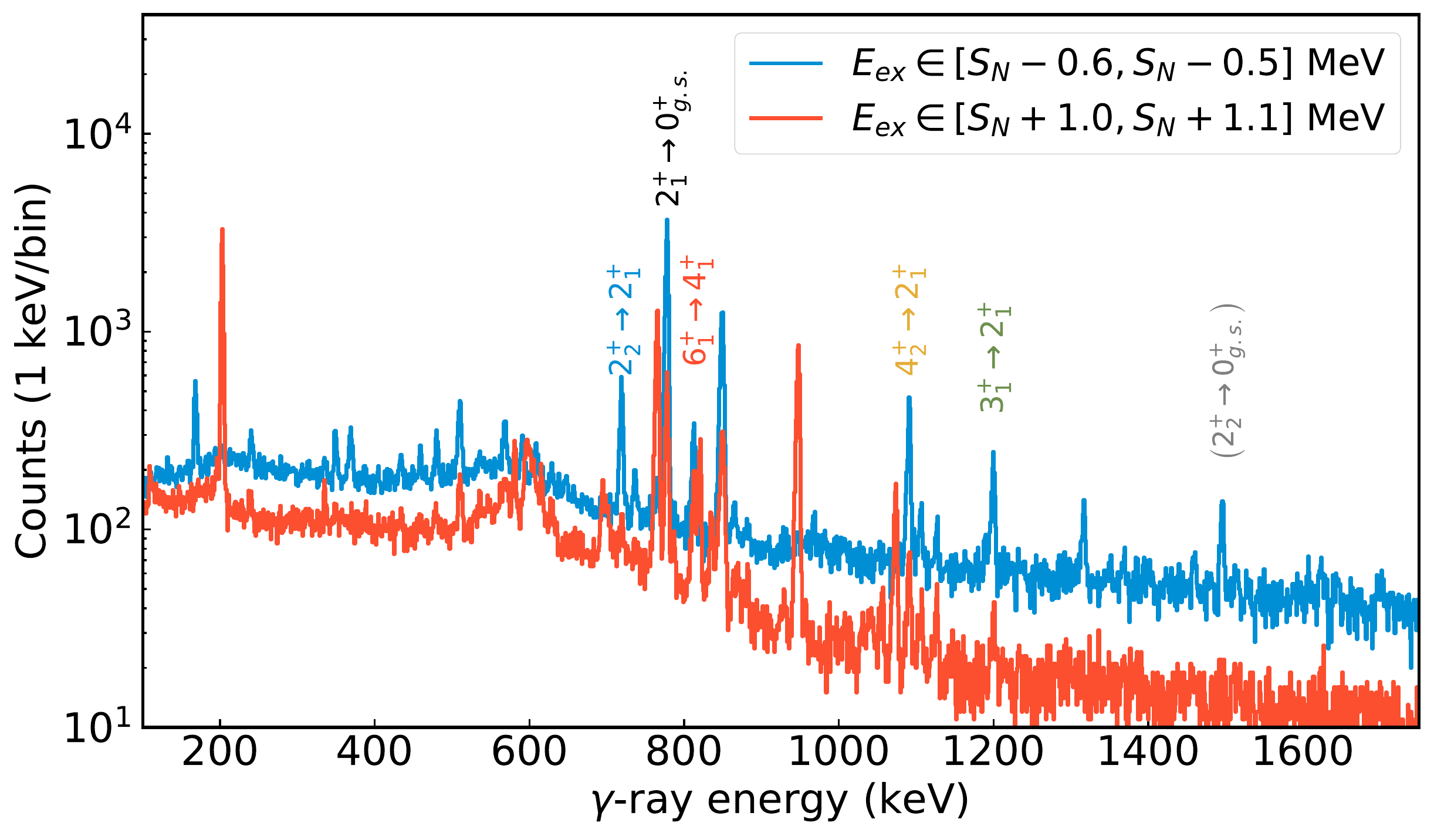}
	\caption{(Color online). Gamma rays in coincidence with protons corresponding to excitation energies in 100-keV bins below $S_{N}$ (blue, decays in $^{96}$Mo) and above $S_{N}$ (red, decays in $^{95,96}$Mo). $^{96}$Mo transitions used to determine $P_{p\gamma}$ are labeled. The single transition which significantly bypasses the $2^{+}_{1}$ state is indicated in parentheses. The persistence of $^{96}$Mo transitions at $E_{ex}\sim$1 MeV above $S_{N}$ highlights the competition between $\gamma$-ray and neutron emission in the CN.	 \label{fig:gspec}}
\end{figure}

In the case of a $(d,p\gamma)$ reaction, the coincidence probability (Eq.~\ref{eqn:full_g_prob}) can be measured as:
\begin{equation}\label{eqn:96mo_g_prob}
P_{p\gamma}(E_{ex}) = N_{p\gamma}(E_{ex})/(N_{p}(E_{ex})\epsilon_{\gamma}).
\end{equation}
Here, $N_{p}$ is the number of detected $(d,p)$ protons, $\epsilon_{\gamma}$ is the $\gamma$-ray photopeak detection efficiency, and $N_{p\gamma}$ is the number of coincidences between a proton and a $\gamma$ ray from the decay of the CN (Fig.~\ref{fig:gspec}). Escher's approach uses Bayesian fits to the experimentally-extracted $P_{p\gamma}(E_{ex})$ (Eq.~\ref{eqn:96mo_g_prob}) to constrain standard expressions for the NLD and $\gamma$SF, which, with $F^{CN}$, are used to determine $G^{CN}_{\chi}(E_{ex},J,\pi)$. The experimentally-constrained parameters are used as inputs for an HF model which is subsequently used to calculate the $(n,\gamma)$ cross section (Eq.~\ref{eqn:HF_ng_CX}) (for details see~\cite{refId0,escherTR738195}). The SRM is thus unique among indirect techniques for determining $(n,\gamma)$ cross sections in that it provides an experimentally-constrained cross section without relying on auxiliary data such as the average radiative width ($\langle \Gamma_{\gamma}\rangle$) or average \textit{s}-wave neutron spacing ($D_{0}$), which are unavailable for exotic nuclei~\cite{refId0,escherTR738195}. 

The $(d,p\gamma)$ reaction was measured in regular kinematics using enriched (98.6\%) $^{95}$Mo targets and a 12.4-MeV deuteron beam produced by the Cyclotron Institute on the College Station campus of Texas A\&M University. The beam had an average intensity of $\sim$0.3 nA and impinged on a 0.96-mg/cm$^{2}$ $^{95}$Mo target. The reaction protons and coincident $\gamma$ rays were measured with the \textbf{S}ilicon \textbf{T}elescope \textbf{A}rray for \textbf{R}eactions with \textbf{Li}vermore, \textbf{Te}xas A\&M, \textbf{R}ichmond (STARLiTeR) apparatus~\cite{Lesher2010286,PhysRevC.90.034601}. The energies of the light-ion ejectiles were measured by a silicon detector telescope located 2.1 cm downstream of the target. The telescope was composed of a thin detector ($\Delta E$, 140 $\mu$m thick) for measuring energy loss and a thick detector ($E$, 1000 $\mu$m thick) to stop protons with energies up to $\sim$18 MeV. Each detector was electronically segmented; the angular resolution was $\sim$1$^{\circ}$. The target chamber was surrounded by four Compton-suppressed high-purity germanium (HPGe) ``clover'' detectors. The trigger condition for data acquisition required that both the $\Delta E$ and E detectors detect a signal above a $\sim$400-keV threshold. When this trigger condition was satisfied, particle detectors and any coincident HPGe detectors were read out. The intrinsic energy resolution of the silicon detectors was determined to be $\sim$20 keV through calibration with an $^{226}$Ra source and the in-beam energy resolution was measured as $\sim$60 keV.


\begin{figure}
\includegraphics[width=8cm,clip]{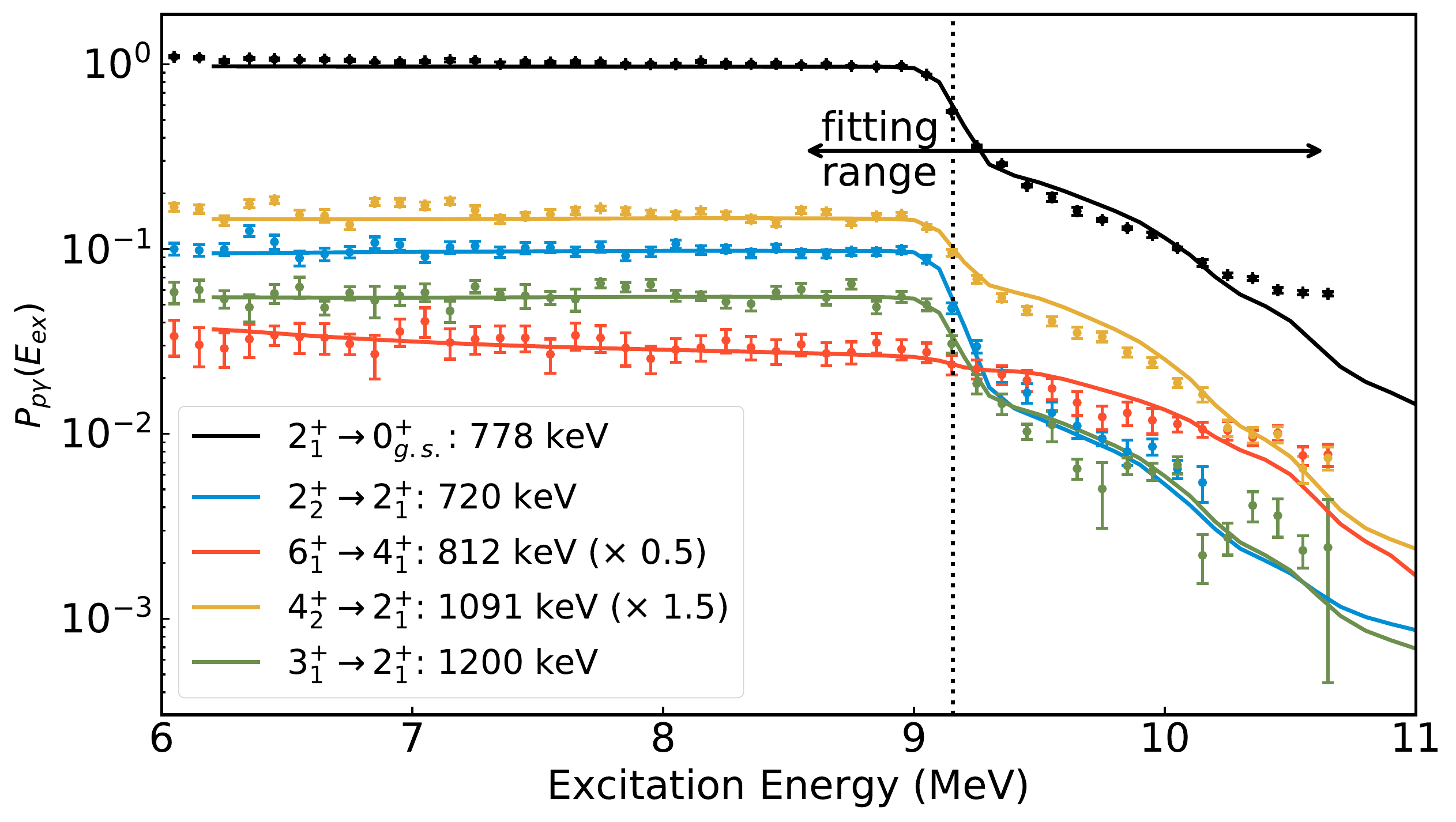}
	\caption{(Color online). Gamma-ray emission probabilities as a function of excitation energy ($P_{p\gamma}(E_{ex})$) for $\gamma$ rays emitted in the decay of excited states in $^{96}$Mo. The vertical dashed line corresponds to $S_{N}$. Data points are experimentally-determined $P_{p\gamma}$. Solid lines represent Bayesian fits to the emission probabilities (same color as the data for each transition). The agreement between data and fits extends well beyond the fitting range.\label{fig:emis_prob}}
\end{figure}

The probability ($P_{p\gamma}$) that a CN formed in a state with energy $E_{ex}$ in the $^{95}$Mo$(d,p\gamma)$ reaction subsequently decays via $\gamma$-ray emission was determined via Eq.~\ref{eqn:96mo_g_prob}. At 12.4 MeV this reaction populates states above and below $S_{N} = 9.15432(5)$ MeV~\cite{Abriola20082501} ($\gamma$ rays in $^{96}$Mo) and above $S_{N}$ ($\gamma$ rays in $^{96}$Mo and in $^{95}$Mo above the 204-keV threshold). Emission probabilities for discrete $\gamma$ rays emitted from the decay of the CN ($^{96}$Mo) were extracted as a function of excitation energy by analyzing particle-$\gamma$ coincidences in 100-keV increments of $E_{ex}$. Gamma-ray yields were obtained from these spectra (Fig.~\ref{fig:gspec}) by Gaussian fits to the photopeaks. Probabilities for several transitions in $^{96}$Mo are shown in Fig.~\ref{fig:emis_prob} as a function of the $E_{ex}$ of the CN. The $2^{+}_{1}\rightarrow0^{+}_{g.s.}$ transition is indeed a strongly collecting transition representing, almost unit probability. In cases where there is not a strong collecting transition, more detailed modeling of the $\gamma$ cascade would be required. This could lead to additional uncertainty, which could, however, be reduced by fitting simultaneously to multiple $\gamma$ transitions (as we have done in this work), measuring the total $\gamma$-ray emission spectrum, or obtaining more structure information.

\begin{figure}
\includegraphics[width=8cm,clip]{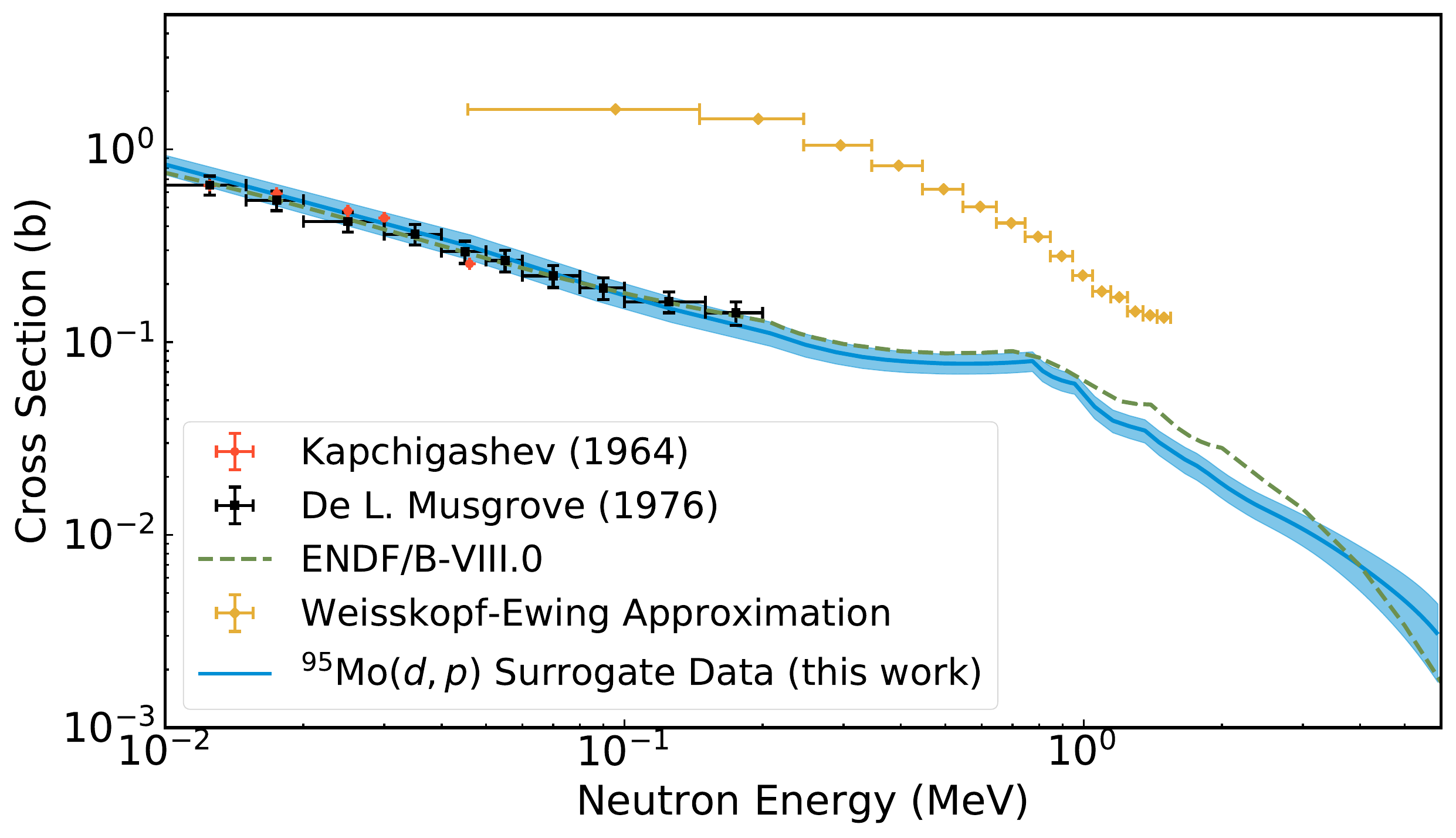}
	\caption{(Color online). Cross sections for the $^{95}$Mo$(n,\gamma)$ reaction. The $(n,\gamma)$ cross section obtained from the SRM (solid blue curve) is in excellent agreement with direct measurements of the cross section~\cite{Kapchigashev1964,DELMUSGROVE1976108} (red circles and black squares). The uncertainty due to experimental data and fitting error is indicated by the shaded band. The result obtained using the WE approximation is also shown (gold diamonds). \label{fig:surrogate_cx}}
\end{figure}

Several experimentally-measured $\gamma$-ray emission probabilities ($P_{p\gamma}$) were fit simultaneously to constrain the decay of the CN. The decay model contained a Gilbert-Cameron level density and a $\gamma$-ray strength function with an energy- and temperature-dependent Lorentzian for the E1 component and an M1 contribution of the Lorentzian shape~\cite{CAPOTE20093107}. The parameters in these functions were adjusted via a Bayesian fit to the data, whose prior encompasses the literature results for this mass region~\cite{CAPOTE20093107}. The fitting results are shown in Fig.~\ref{fig:emis_prob}, with the measured surrogate coincidence probabilities $P_{p\gamma}$ displayed with statistical uncertainties. The agreement between the data and the fits is excellent over the fitting range used, both above and below $S_{N}$. These constrained HF-model parameters were then used with Eq.~\ref{eqn:HF_ng_CX} to deduce the cross section for $^{95}$Mo$(n,\gamma)$, shown in Fig.~\ref{fig:surrogate_cx}. The $\sigma^{CN}$ values were calculated using the neutron-nucleus optical potential parameters from~\cite{KONING2003231}. The resulting cross section is shown with an uncertainty band that arises from the experimental uncertainties and the error in the Bayesian fit. Uncertainties arising from the choice of the deuteron and neutron optical potentials are expected to be negligible for this case, as they have been found to have little impact on the entry spin distribution. For applications away from stability, where the optical potentials are less well known, this has to be revisited. However, the simultaneous measurement of the angular distributions of reaction protons and elastic scattering from a $(d,p)$ measurement in inverse kinematics could constrain the optical potentials. Overall, the present result is in excellent agreement with previous direct measurements of the $^{95}$Mo$(n,\gamma)$ cross section~\cite{Kapchigashev1964,DELMUSGROVE1976108} and to the cross section reported in the ENDF/B-VIII.0 evaluation~\cite{BROWN20181}.


To demonstrate the importance of the proper treatment of the spin-parity distribution produced in the surrogate reaction (shown in Fig.~\ref{fig:greg_calc}), we also show (Fig.~\ref{fig:surrogate_cx}) the cross section obtained when the WE approximation is employed. Obviously it is not appropriate to employ the WE approximation when determining $(n,\gamma)$ cross sections from the $(d,p)$ data. This over-estimation of the $(n,\gamma)$ cross section was also observed in previous studies employing the WE approximation~\cite{PhysRevC.81.034608,BOUTOUX2012319,PhysRevC.81.011602}. The current work confirms previous suggestions (cf. ~\cite{PhysRevC.81.034608,BOUTOUX2012319,PhysRevC.81.011602,PhysRevC.75.055807,RevModPhys.84.353,PhysRevC.81.024612,PhysRevC.92.034611}) that a proper treatment of such differences is critical to accurately constrain the $(n,\gamma)$ reaction cross section through the SRM. 

In summary, we have demonstrated that a measurement of the $(d,p)$ reaction, when combined with the proper theoretical treatment, can be used to indirectly determine $(n,\gamma)$ cross sections. The $^{95}$Mo$(d,p\gamma)$ reaction was measured to validate the $(d,p\gamma)$ reaction as a surrogate for neutron capture, a reaction important for the synthesis of almost all of the elements heavier than iron and for applications in nuclear energy and security. This Letter shows the power of the SRM developed in Refs.~\cite{RevModPhys.84.353,refId0,escherTR738195} with the proper treatment of the spin-parity distribution ~\cite{PhysRevC.92.034611,Potel2017} of the CN created in $(d,p)$. This approach moves beyond the WE approximation which has been previously shown, and here confirmed, to be inadequate for neutron capture~\cite{PhysRevC.81.034608,BOUTOUX2012319,PhysRevC.81.011602}. We show that a robust model of the formation of the CN~\cite{PhysRevC.92.034611,Potel2017} and proper treatment of its decay~\cite{refId0,escherTR738195} within the framework of the SRM~\cite{RevModPhys.84.353,refId0,escherTR738195} is necessary to extract from $(d,p\gamma)$ data a capture cross section that agrees with the directly-measured $(n,\gamma)$ cross section. We note that the kinematics of the $(d,p)$ reaction are ideal for measurements with short-lived beams in inverse kinematics. Therefore, the $(d,p\gamma)$ surrogate reaction is a promising tool to extract $(n,\gamma)$ reaction cross sections for exotic, \textit{r}-process nuclei and for nuclei created in other high-neutron-fluence environments. The benchmarking of the Surrogate Reactions Method for $(n,\gamma)$ with measurements of the $(d,p\gamma)$ reaction presented here opens the door to important measurements in an exciting area of the nuclear chart which is becoming increasingly accessible at modern accelerator facilities.

\begin{acknowledgments}
The authors thank the staff of the Texas A\&M Cyclotron Institute for providing the beams used in this work. This work was supported in part by the U.S. Department of Energy National Nuclear Security Administration under the Stewardship Science Academic Alliances program, NNSA Grants No. DE-FG52-09NA29467 and No. DE-NA0000979, Lawrence Livermore National Laboratory Contract No. DE-AC52-07NA27344 and LDRD 16-ERD-022, Texas A\&M Nuclear Physics Grant No. DE-FG02-93ER40773, the Office of Nuclear Physics, and the National Science Foundation. 

\end{acknowledgments}

\bibliography{ratkiewicz_biblio}
\end{document}